\documentclass[12pt,preprint]{aastex}

\newcommand{\vvec}{{\bf v}}

\newcommand{\beq}{\begin{equation}}
\newcommand{\eeq}{\end{equation}}
\newcommand{\We}{{W\!e}}
\newcommand{\Rey}{{Re}}
\newcommand{\B}{{\bf B}}

\received{}
\revised{}
\accepted{}
\ccc{}
\cpright{}{}

\shorttitle{Viscoelastic Jets}

\begin{document}
\title{Viscoelastic Analogy for the Acceleration and Collimation of Astrophysical Jets}
\author{Peter T. Williams}
\affil{Department of Physics, University of Florida, Gainesville, FL 32611 \altaffilmark{1}}
\altaffiltext{1}{Present address: PMB\# 225, 3233 N. Mesa, Suite 205, El Paso, TX 79902}
\email{peter@phys.ufl.edu}

\begin{abstract}
Jets are ubiquitous in astronomy. 
It has been conjectured that the existence of jets is intimately
connected with the spin of the central object and the viscous angular momentum transport of the inner disk. 
Bipolar jet-like structures propelled by the viscous torque on a spinning central object are also known in a
completely different context, namely the flow in the laboratory of a viscoelastic fluid. 
On the basis of an analogy of the tangled magnetic field lines of magnetohydrodynamic (MHD) turbulence to the tangled polymers of 
viscoelastic polymer solutions, we propose a viscoelastic description of the dynamics of highly turbulent conductive fluid. We argue
that the same mechanism that forms jets in viscoelastic fluids in the laboratory 
 may be responsible for collimating and powering astrophysical jets by the angular momentum of the
central object.
\end{abstract}
\keywords{MHD --- turbulence --- stars: winds, outflows --- galaxies: jets}

\notetoeditor{There are some excellent photos of the laboratory phenomenon that would be nice to include: Fig 4 and Fig 5 from
Giesekus (1963). This is published by John Wiley & Sons, LOC #50-19776, no ISBN given. Alternatively there is a nice sequence
of photos in fig 15 on p. 25 of Illustrated Experiments in Fluid Mechanics, published by the MIT Press, Cambridge, 1972,
ISBN 0-262-14014-4. I have contacted them and they say the copyright is owned by Educational Development Center, Inc.
Better quality photos are also available in
Giesekus (1994) but I do not have access to that book at the present time. A photo from any of these sources would be 
very nice to include if that is possible.
I do not know what your policy is in this regard. Thank you.}

\section{Introduction}
Viscoelastic behavior
is characteristic of fluids such as polymer suspensions in which long molecular chains become entangled. The deformation
and disruption of this entanglement that occurs when the fluid is subjected to shear, and the relaxation to a new entangled state,
give such fluids a ``memory'' that ordinary Newtonian fluids do not have. Egg whites are one common household example.
These fluids are in a sense midway between being 
elastic solids and Newtonian
fluids, being more one or the other depending upon the ratio of the relaxation time of the
entangled polymers to the deformation time scale. This ratio is known as the Weissenberg number $\We$.
Viscoelastic behavior can cause flows that are qualitatively quite different from the Newtonian case,
such as the Weissenberg effect, in which a fluid climbs a spinning rod.

We propose that the effects of ``mesoscale'' magnetic fields in some instances of MHD turbulence can be modeled by additional
viscous terms, and that the structure of these terms is similar to that of a viscoelastic fluid. (It has come to our attention that 
this has also
been conjectured in a recent paper by Ogilvie \cite{Ogil:2001}, who demonstrated that a viscoelastic model of MHD turbulence
can dramatically affect the dynamics of eccentric accretion disks.) 
It has long been noted that the dynamics of vortex filaments of fully-developed turbulence bear a superficial
likeness to a network of polymers; see 
Chorin (1988).
There is also an obvious analogy between the tangled polymers of a polymer fluid and the tangled field lines
in MHD turbulence in the case that most of the field energy is on intermediate to small scales. 
However, in contrast with polymers which are typically conserved, the magnetic energy on small scales is not conserved and may dissipate.
We therefore restrict our attention to flows in which
the rate of shear changes slowly with respect to the dissipation time scale.
We assume that the large-scale
({\it i.e.} mean-field) magnetic field is rather weak, but the magnetic field on intermediate scales is advected by hydrodynamic
turbulence \cite{Ott:1998}. The particular case we consider is turbulence driven by shear, such as in a Balbus-Hawley magnetic
shearing instability \cite{BalHaw:1998} in the case that the net magnetic flux is negligible.

One of the more startling effects seen in viscoelastic fluids is the reversal of the secondary flow around a rotating sphere
at moderate rotation speeds.
A Newtonian fluid will flow inwards along the rotation axis and be expelled in the equatorial plane, but in a viscoelastic
fluid, the secondary flow 
is exactly the opposite. We propose this phenomenon as a model for the production of astrophysical jets powered in part
by the angular momentum of the central object. 

\section{MHD ``rheology''}

In MHD, the magnetic field evolves according to the vector advection-diffusion equation
\beq
\partial_t \B = \B\cdot \nabla \vvec - \vvec \cdot \nabla \B - \B (\nabla \cdot \vvec) + \eta \nabla^2 \B,
\eeq
where $\B$ is the magnetic field, $\vvec$ is the fluid velocity and $\eta$ is the magnetic diffusivity, which is determined by the conductivity of the fluid.
If the diffusivity $\eta$ is zero, the flux-freezing is perfect.
The contribution of magnetic fields to the stress tensor $\tau_{ij}$ is given by
\beq
\tau_{ij} = - \left( \B^2 \over 8 \pi \right) \delta_{ij} + {1 \over 4 \pi} B_i B_j.
\eeq
Let us average spatially over some ``subgrid'' length scale, and concentrate on the curvature term.
Under perfect flux-freezing, the tensor $M_{ij} = \langle B_i B_j \rangle$ can be shown from eq. (1) to evolve according to
\beq
D_t(M_{ij}) - M_{ik}L_{kj} - L^T_{ik}M_{kj} + 2 M_{ij} L_{kk} = 0  
\eeq
where $L_{ij} = \partial_i v_j$ and $D_t$ is the usual advective derivative. The first three terms in
eq. (3) are known collectively in rheology as the upper-convected invariant derivative ${\cal D}^{[u]}_t$
of the tensor $M_{ij}$. Insofar as the viscous stress in a non-Newtonian fluid does not depend only
on the velocity gradient $L_{ij}$, but on its history as well as higher-order derivatives of the
velocity, the concept of the advective derivative $\partial_t + \vvec \cdot \nabla$ must be extended
to respect the material frame indifference of the stress tensor.  
The upper-convected invariant derivative ${\cal D}^{[u]}_t$ of the viscous stress tensor is 
a common material derivative in rheology, and it is used in many rheological models.

According to flux-freezing, as expressed in eq. (3),
on- and off-diagonal components of the magnetic stress in a  highly conducting plasma threaded with a small-scale 
magnetic field will grow in response to large-scale shear. 
A simple analogy is a square piece of rubber on which we have drawn a
tangled, statistically homogeneous and isotropic scribble. If we pull the upper edge of this sheet to the right, say, and the lower edge of the
sheet to the left, the area of the square will remain the same, but the scribble is now no longer isotropic. There is an overall tendency
of the scribble to be composed of lines oriented diagonally from the lower left corner to the upper right corner.

With this picture in mind, suppose that at some time $t=0$ the individual components of the magnetic field are distributed as Gaussian random variables
with equal variance $\sigma$ and zero mean. In a Cartesian reference frame,
given a uniform shear $\gamma$ in the $\hat x$ direction, with perfect flux-freezing, the mean field will evolve according to
\beq
\langle M_{ij} \rangle = \sigma^2 \left( \matrix{
1 + (\gamma t)^2 & \gamma t & 0 \cr
\gamma t & 1 & 0 \cr
0 & 0 & 1 \cr
} \right).
\eeq
The field lines are in reality constantly shifting and being arranged on small scales by 
small turbulent eddies, which we characterize by a relaxation time $s$. This time incorporates the effects of the turbulent stretching and
twisting that generates the field, as well as the dissipative processes that are governed by the magnetic diffusivity $\eta$.
In the case we consider, the magnetic field draws its energy from the shear, so 
it is natural to assume that $1/s$ is comparable to the rate of shear $\gamma$.
We therefore expect that the product 
$\gamma s$, which is the Weissenberg number $(\We)$, is of the order of magnitude of one or greater.
(Ogilvie finds $\We \sim 1$ based on a comparison with simulations.)
If the turbulence is driven by other processes, such as convection, the effective Weissenberg number might be very different from $1$.

In the case of steady shear, we approximate the form of the tensor $\langle B_i B_j \rangle$ by replacing $t$ in the above
expression with the relaxation time $s$. The off-diagonal stress is proportional to $\gamma s = (\We)$, with the constant of
proportionality determined by the viscosity prescription. There is a normal stress difference in the direction of
shear $\tau_{xx} - \tau_{yy} = (\We) \tau_{xy}$. This form of the stress tensor, modulo the magnetic pressure,
 is very similar to the second-order approximation
to a viscoelastic Weissenberg fluid, with zero second normal stress difference $\tau_{yy} - \tau_{zz}$. The non-Newtonian viscous
stress in such a fluid may be written in terms of the Rivlin-Erikson tensors $A^{[1]}$ and $A^{[2]}$
as
$\tau = \mu A^{[1]} + \alpha_1 A^{[2]} + \alpha_2 (A^{[2]})^2$ where $\mu$ is the usual viscosity
and the second-order coefficients $\alpha_1$ and $\alpha_2$ satisfy $\alpha_2 = - 2 \alpha_1$.
The first Rivlin-Erikson tensor $A^{[1]}$ is simply the symmetrized velocity gradient,
$A^{[1]}_{ij} = L_{ij} + L_{ji}$, and subsequent tensors $A^{[n]}$ obey the recursion relation
$A^{[n+1]}_{ij} = D_t A^{[n]} + A^{[n]}_{ik}L_{kj} + A^{[n]}_{jk}L_{ki}$. The $n$th Rivlin-Erikson
tensor gives the $n$th time derivative of the length of an infinitesimal material line $|d{\bf x}|$ embedded in the fluid,
so that $d_t^{(n)} |d{\bf x}|^2 = A^{[n]}_{ij} dx_i dx_j$. A second-order fluid is thus simply an approximation
to a non-Newtonian fluid that is second-order in the velocity gradient, having contributions to the
viscous stress that are proportional to the second spatial derivative of the velocity as well as
the square of the first spatial derivative of the velocity, in addition to the normal Newtonian viscous
stress which is proportional to the (symmetrized) first spatial derivative of the velocity.
As has been noted by B\"ohme (1987), the term ``second-order fluid'' is thus somewhat unfortunate since
it is the flow, and not the fluid, which is really second-order.
If the material derivative of the rate-of-strain tensor itself changes on a time scale
comparable or faster than $s$ then a more sophisticated model must be brought to bear. 

In a cylindrical coordinate system with azimuthal shear,
such as in the fluid surrounding a rotating sphere, the normal stress difference contributes to a global hoop-stress $\tau_{\theta \theta}$,
which acts like a rubber-band, 
creating a radially-inward ``non-Newtonian'' specific force $f_{\rm nn} = -\tau_{\theta \theta} / (r \rho)$.
In the laboratory, the hoop-stress-induced forces cause a radially-inward flow close to the equatorial plane  \cite{Gies:1963,Gies:1994}. 
This has also been shown mathematically up to various orders in $\We$ and $\Rey$ for a second-order fluid \cite{Gies:1963, Bohm:1987, Jose:1990, Gies:1994} as well as to leading order for a viscoelastic Oldroyd model B' fluid
\cite{ThoWal:1964} which has a spectrum of relaxation times. 
This inflow pushes and collimates the intervening material into bipolar
axial outflows for moderate $\We$ and Reynolds number $\Rey$. 

\section{Jet model}

The astrophysical analog we propose is a spinning massive central object immersed in a viscous thick disk
with a tangled magnetic field.
The solutions in the rheology literature do not map directly onto astrophysical solutions because gravitational, compressibility,
and radiative effects are important
in the latter. Furthermore, the viscosity in the astrophysical case is probably not constant as we 
assume here. We therefore confine ourselves to general dimensional arguments.
We will allow the central object to have an extended magnetosphere that acts like a rigid rotator, so that 
the effective radius of the object $R_0$ may in fact be much larger than the actual radius of the object,
depending upon the strength of the magnetosphere of the object. 
We assume that the disk and central object are viscously coupled so that $\Omega_0 = \Omega(R_0)$, and
we define $\tilde \omega$ to be the angular speed of the central object
$\Omega_0$ normalized to the Keplerian frequency at the object's effective surface $\Omega_K(R_0)$.
Note that we may have $\tilde \omega > 1$. 
The central object rotates
in the same direction as the disk. Also we define $r = R / R_0$.
Define $r_m$ to be the dimensionless radius of the thick disk, and let the angular frequency of the disk
$\Omega$ scale as $R^{-q}$. Our conclusions are not sensitive to this last assumption, but it provides a nice framework to discuss
the viscoelastic effects. 
Note that if we take $q > 3/2$ for $\tilde \omega > 1$ and $q<3/2$ for $\tilde \omega < 1$, the
 thick disk will match to a Keplerian disk at $r^{q-3/2} = \tilde \omega$

The structure of thick disks is sensitive to the viscosity prescription, and the ``correct'' viscosity is unknown.
We assume that $R_0 < H < r_m R_0$ where $H$ is the thickness of the thick disk.
As an upper and lower bound, we set
\beq
\nu = \alpha \beta R_0^2  \left( {\partial \Omega \over \partial \ln R } \right)_0.
\eeq
where $\beta$ is between $1$ and $r_m^2$
 and $\alpha \sim 1$ as usual. The traditional parameter $\alpha$ is thus redundant here. 
The effective Reynolds number as determined by $R_0$ and $\Omega_0$ is then $\Rey =  1/(q \alpha \beta)$, which is on the
order of $1$ or less.

The viscous stress is $\tau_{r \theta} = \rho \nu (R \Omega')$. The specific 
inward ``non-Newtonian'' radial force $f_{nn}$ 
due to the normal stress difference is then
\beq
f_{nn} = - (\We) \nu |\Omega'| = - (\We) q^2 \alpha \beta R_0^2 \left({ \Omega^2 \over R } \right).
\eeq
It must not be forgotten that the viscous and hoop stresses occur in proportion to the magnetic pressure.
From eqs. (2) and (4), and recalling that $\tau_{r\theta} = \nu (R \Omega')$, we find
\beq
{1 \over \rho} P_{\rm mag} = {1 \over 8 \pi} \nu |R \Omega'| \left( {3  \We^{-1}} + \We \right) = 
{C_1 \over 8 \pi} \nu | R \Omega' |.
\eeq
The specific 
centrifugal inertial force is $f_{\rm inert} = \Omega^2 R$ and the specific
gravitational force is $f_{\rm grav} = -\Omega_K^2 R$, so that
the inward hoop stresses will dominate inertial forces for $r^2 < (\We) q^2 \alpha \beta$, and they will dominate gravitational
forces for $r^{(2q-1)} < (\We) q^2 \alpha \beta \tilde \omega^2$. 
However, it is important to 
note that, in the incompressible case, or for a polytrope, the specific forces
 $f_{\rm grav}$, $f_{\rm inert}$, and $f_{\rm press}$ are {\em all derivable from potentials},
whereas $f_{\rm nn}$ is not. Note that 
the specific pressure force $f_{\rm press} = \rho^{-1} \nabla P$ may be written as the gradient of
some function so long as there is a one-to-one relation between $P$ and $\rho$.
We expect that there is typically no realistic
combination of gravitational, centrifugal, or pressure forces that
will balance the non-Newtonian
 hoop stresses for an arbitrary choice of $v_{\theta}(R,z)$, regardless of the value of $\beta$.
For the choice of $\partial_z v_{\theta} = 0$,
such as in a thin disk, the hoop stress $f_{\rm nn}$ depends only on $R$ and thus may be written $f_{\rm nn} = \nabla \Phi$ for some
$\Phi$, but for a thick disk this will not typically be the case. 

In contrast with azimuthal viscous forces which dissipate kinetic
energy, $\bar f_{\rm nn} \cdot \bar v$ is positive for a radial inflow, so that the hoop stresses do positive work
on the fluid. There are no hoop stresses along the rotation axis of the system, so the energy gained by the fluid is free
to escape in this direction. Lastly, as in the laboratory case, it is hypothesized that these hoop stresses 
naturally confine and collimate the flow, if the thick disk is sufficiently larger than the object radius $R_0$.

The hoop stresses are powered in part by the mechanical accretion luminosity, and the specific work performed by them for a fluid element
in the equatorial plane that moves from the outer radius $R_m$ to $R_0$ is
\beq
W = \int_{R_0}^{R_m} f_{\rm nn} \ dR = {1\over 2} q \alpha \beta (\We) {G M \over R_0} (1 - r_m^{-2q}) 
\eeq
For $\beta \gg 1$ this work can be much larger than the binding energy; some of this energy will be
viscously dissipated, and some is work against the pressure gradient.

 Extrapolating from the laboratory case for moderate $\We$ and $\Rey$,
the launch speed  of the jet is expected to be of the order of
\beq
v_{\rm jet} = C_2 {\nu \over R_0} \left({ \Omega_0 R_0^2 \over \nu } \right)^2 = C_2 {R_0 \Omega_0 \over q \alpha \beta},
\eeq
where $C_2$ is a constant of order unity.
In escaping to infinity, the jet will lose its binding energy $GM/R_0 = (R_0 \Omega_0 / \tilde \omega)^2$, but
it will be propelled by the magnetic pressure $P_{\rm mag}/\rho$ given in eq. (7). 
If the thick disk is substantially thicker than the object
effecive radius $R_0$ then the jet may also suffer some viscous drag, which we estimate by Poiseuille's law.
In traversing the thickness of the disk $H$ a jet of radius $aR_0$ (where we expect $a \sim 1$) will lose an amount
of energy per unit mass of the order of $\nu v_{\rm jet} H /(a R_0)^2 = C_3 H R_0 \Omega_0^2$ where $C_3$
is of order unity.
We thus very roughly estimate the jet speed at infinity as
\beq
{1\over 2} v_\infty^2 \sim R_0^2 \Omega_0^2 \left(
{C_1 q^2 \alpha \beta \over 8 \pi} +
{C_2\over 2(q\alpha \beta)^2} -
{C_3 H \over R_0}
 - {1 \over \tilde \omega}
\right).
\eeq 
The speed of the jet naturally scales as $R_0 \Omega_0$, and increases with the speed of rotation $\tilde \omega$, which
must be close to breakup for $\beta \sim 1$ for the jet to reach infinity.  
We caution however that further work is required to find to what degree the above admittedly very
approximate relationship
holds in the astrophysical case. This expression is intended only to show roughly the relative importance of the
various effects we have considered.
We have not considered any relativistic effects, and we have not addressed the importance of the accretion
efficiency, nor any effects associated with the central object having an event horizon. Accretion onto
a black hole will certainly complicate matters very much. As it stands, therefore, we believe
our present arguments apply perhaps best to the case of jets from protostellar objects.
\section{Discussion}
We have proposed that the fluid dynamics of a highly-conductive turbulent medium with tangled magentic fields is similar
in some respects to the viscoelastic fluid dynamics of a suspension of tangled polymers. We discuss the production of
jet-like structures in viscoelastic fluids induced by a spinning sphere, and outline an analogous model for the 
production of astrophysical jets propelled by a spinning central object.
The processes outlined above are thus hypothesized to propel jets by tangled magnetic fields,
as has been previously suggested by Heinz \& Begelman (2000).

We believe that this tentative model for the acceleration of jets has some advantages over previous models.
Firstly, the jet process does not rely upon the insertion ``by hand'' of any large-scale magnetic field. 
The jet also appears to be self-collimating, and it is a solid-core jet, rather than a hollow-core jet. As pointed
out by 
Li (2001),
 tangled magnetic fields may also stabilize the jet from the long-wave mode kink instability that plagues standard 
MHD jet models.  Lastly, the jet mechanism
 we describe is powered in part by the angular momentum of the central object.
We hope to improve upon this work with more detailed calculations in a future paper.

\acknowledgments
{\it
The majority of this work was completed when the author was a graduate
student at the University of Texas at Austin.
The author wishes to thank Craig Wheeler for helpful comments,
John Scalo and Nairn Baliber for encouraging
him over the years to write this up,
and Robert Buchler for giving
him the time and financial support necessary to do so.
}



\end{document}